\documentclass[
 reprint,
nofootinbib,
 amsmath,amssymb,
 aps,
]{revtex4-2}

\usepackage[dvipdfmx]{graphicx}
\usepackage{dcolumn}
\usepackage{bm}
\usepackage{color}
\usepackage[colorlinks=true, linkcolor=blue, citecolor=red, urlcolor=blue]{hyperref}
\usepackage{physics}
\usepackage{orcidlink}
\usepackage{tikz}
\usetikzlibrary{positioning, shapes.geometric}
\usetikzlibrary{arrows.meta}
\usetikzlibrary{calc}
\usepackage{xspace}

\newcommand{\nd}{n_{\rm d}}
\newcommand{\ns}{n_{\rm s}}
\newcommand{\np}{n_{\rm p}}

\newcommand{\sign}[1]{{\rm sign}\!\left(#1\right)}
\newcommand{\dmoped}{{\it d}-MOPED\xspace}
\newcommand{\pmoped}{{\it p}-MOPED\xspace}
\renewcommand{\tr}{{\rm Tr}}

\begin{document}

\title{
Data Compression with Noise Suppression for Inference under Noisy Covariance
}

\author{Sunao Sugiyama\orcidlink{0000-0003-1153-6735}}\email{ssunao@sas.upenn.edu}
\author{Minsu Park\orcidlink{0000-0001-5972-4191}}
\affiliation{Department of Physics and Astronomy, University of Pennsylvania, Philadelphia, PA 19104, USA}

\date{\today}

\begin{abstract}
In many fields including cosmology, statistical inference often relies on Gaussian likelihoods whose covariance matrices are estimated from a finite number of simulations. This finite-sample estimation introduces noise into the covariance, which propagates to parameter estimates, a phenomenon known as the Dodelson-Schneider (DS) effect, leading to inflated uncertainties. While the Massively Optimized Parameter Estimation and Data compression (MOPED) algorithm offers lossless Fisher information-preserving compression, it does not mitigate the DS effect when the compression matrix itself is derived from noisy covariances. In this paper, we propose a modified compression scheme, powered MOPED ($p$-MOPED), which suppresses noise propagation by balancing information retention and covariance estimate noise reduction through a tunable power-law transformation of the sample correlation matrix. We test $p$-MOPED against standard and diagonal MOPED on toy models and on cosmological data from the Subaru Hyper Suprime-Cam Year 3 weak lensing survey. Our results demonstrate that $p$-MOPED consistently outperforms other approaches, especially in regimes with limited simulations, offering a robust compression strategy for high-dimensional data analyses under practical constraints.
\end{abstract}

\maketitle

\section{Introduction}
In many modern cosmological analyses, particularly those involving weak gravitational lensing or large-scale structure, high-dimensional data vectors are used to extract maximal information from surveys. These include two-point and higher-order correlation functions measured across multiple redshift and angular bins
(see e.g. \citep{Li.Wang.2023,Dalal.Wang.2023,Secco.To.2021,Wright.Zhang.2025zw} and \citep{Philcox.Ivanov.2021,Burger.Martinet.2023,Gomes.Weller.2025} for two-point and higher-order correlation analyses applied to the existing galaxy survey data sets). 
When multiple observables are combined to break parameter degeneracies\citep[see e.g.][from Stage-III imaging survey]{Sugiyama.Wang.2023,Collaboration.Zuntz.2021}, the dimensionality of the data vector can easily exceed several hundred or even more. This is further exacerbated by the fact that future surveys will have an increasing number of redshift and scale bins.
In order to derive the reliable parameter estimate and its associated error from such a huge data vector, accurate covariance estimation, as one of the core ingredients of the Gaussian likelihood, is essential \citep{Anderson.Anderson.2003,Hartlap.Schneider.2007,Friedrich.Eifler.2017,Joachimi.Joachimi.2016}. 

In many cases, the covariance matrix must often be estimated from a finite set of mock realizations in order to capture the non-trivial effects such as a non-linear relation of the underlying random field and summary statistics, the survey mask effect, etc. A major complication arises when the number of simulations $n_{\rm s}$ is not significantly larger than the dimension of the data vector $n_{\rm d}$. Since the covariance matrix contains $n_{\rm d}(n_{\rm d}+1)/2$ degrees of freedom, limited simulations result in a noisy sample covariance estimate and its noisy inverse. This noise propagates through the likelihood function and 
introduces the scatter in the maximum-likelihood parameters
-- a phenomenon known as the Dodelson-Schneider (DS) effect \cite{Dodelson.Schneider.2013}, which must be carefully addressed to ensure the reliability of the analysis \citep[see][for the marginalization of the uncertainty due to noise propagation]{Percival.Weaver.2013}.

The DS effect becomes severe when the data dimension is large compared to the number of simulations $n_{\rm s}$, and therefore we expect the effect can be mitigated by performing the data compression.
The Massively Optimized Parameter Estimation and Data compression (MOPED) method \cite{Heavens.Lahav.1999} is a widely-used linear compression technique that preserves Fisher information under ideal conditions. However, when the MOPED compression matrix is constructed using the noisy sample covariance, the noise is transmitted to the compressed covariance matrix. As a result, the DS effect is not mitigated by standard MOPED compression. Note that the same problem should arise in other compression methods \cite[][for example]{Alsing.Wandelt.2017,Tegmark.Tegmark.1997}, which rely on the noisy sample covariance to construct compression procedure (the compression matrix in the case of linear compression).

This paper addresses this issue by proposing a modified compression approach designed to suppress noise propagation in the compressed covariance. Rather than strictly preserving Fisher information as in standard MOPED, our method is designed to balance the preservation of information content against the reduction of noise due to finite-sample covariance estimates. This trade-off allows for more stable and accurate parameter inference, particularly in regimes where the number of simulations is not sufficiently large—as is often the case in realistic cosmological analyses.

This paper is organized as follows. In Section~\ref{sec:prerequisite}, we introduce the basic mathematical background on the noise propagation to parameter inference. In Section~\ref{sec:moped-variants}, we introduce a standard linear data compression method called MOPED and variants of them as a proposed method in this paper. In Section~\ref{sec:application}, we apply the methods to a realistic real-world problem, Subaru HSC weak lensing data, as an example of cosmological large-scale structure data. In Section~\ref{sec:summary}, we conclude the paper with outlook.

\section{Prerequisite}\label{sec:prerequisite}
\subsection{Definition of basic quantities}\label{sec:basic}
We consider the problem with the following Gaussian likelihood
\begin{align}
    \ln\mathcal{L}_C
    = -\frac{1}{2}[d-t(p)]^T C^{-1}[d-t(p)] \ ,
    \label{eq:true-like}
\end{align}
where $d$ is a data vector with dimension $\nd$, $t(p)$ is the model prediction for the data with the model parameter $p$ with dimension $\np$, and $C$ is the covariance matrix. We assume that the covariance is parameter-independent. We neglected the constant additive term that corresponds to likelihood normalization. For simplicity, we assume the theoretical model is linear to the model parameter, $t(p)=M p$, where $M=\partial t/\partial p$ is a response matrix of $(\nd, \np)$ shape. 
Here we assume that the model perfectly describe the data i.e., there exists a true parameter $p^{t}$ such that the data are generated according to the model with this parameter. We do not consider any model mis-specification in this work.

When the covariance matrix is not known or difficult to theoretically predict, we usually replace it with the sample covariance which is estimated from the finite number of simulations at a fixed fiducial point in parameter space. Let $d^{(r)}$ to be the $r$-th sample of the data vector that follows the true Gaussian likelihood Eq.~(\ref{eq:true-like}), and we obtain the sample covariance
\begin{align}
    S = \frac{1}{\ns-1}\sum_{r=1}^{\ns}
    \left(d^{(r)}-\bar{d}\right) 
    \left(d^{(r)}-\bar{d}\right)^T \ ,
    \label{eq:S}
\end{align}
where $\bar{d}=\sum_{r=1}^{\ns}d^{(r)}/\ns$ is the sample mean of the data vector, and $\ns$ is the number of samples used to estimate the sample covariance.

\subsection{Dodelson-Schneider effect}
When we replace the true covariance in Eq.~(\ref{eq:true-like}) with the sample covariance to approximate the likelihood by $\mathcal{L}_S$, the noise in the sample covariance propagates to the parameter estimate, causing scatter in the best fit parameter.
This effect is first derived in Dodelson-Schneider \cite{Dodelson.Schneider.2013}, and the associated correction factor that is needed to be multiplied to the sample covariance matrix is called DS factor. Here we review how the DS factor is derived below.

For a given data vector $d$, the maximum likelihood or best-fit parameter for the likelihood with the true covariance $\mathcal{L}_C$ is obtained by
\begin{align}
    \hat{p}(C) 
    &= p^{\rm t} + F_C^{-1}M^T 
    C^{-1}(d-t^{\rm t}) \nonumber \\
    &\equiv p^{\rm t} + X_C (d-t^{\rm t})  \ ,
\end{align}
where $p^{\rm t}$ and $t^{\rm t}=M p^{\rm t}$ are the underlying true parameter and prediction, 
and $F_C$ is the Fisher matrix
\begin{align}
    F_C = M^T C^{-1} M  \ ,
\end{align}
evaluated with the true covariance as indicated by the subscript.
Also we defined the matrix combinations $X_C=F_C^{-1}M^TC^{-1}$.
When we use the sample covariance in the likelihood, then the best-fir parameter is 
\begin{align}
    \hat{p}(S) = p^{\rm t} + 
    X_S(d-t^{\rm t}) \ .
\end{align}
By taking the average with respect to the data realization, we find $\langle\hat{p}(S)\rangle_d=\langle\hat{p}(C)\rangle_d=p^t$, meaning that the expectation value of the best-fit parameter is {\it not} biased due to the use of sample covariance $S$.
However, its variance is inflated by a factor of 
\begin{align}
    f_{\rm DS} = \frac{
    \tr\left( X_S C X_S^{T} \right)}{
    \tr\left( X_C C X_C^{T} \right)}
    \label{eq:ds-factor-tr}
\end{align}
and therefore, the final posterior width need to be inflated by the square root of this factor, or equivalently the sample covariance need to be inflated by this factor during inference, where the latter approach should be more stable for parameter sampling in practice.
The authors of \citet{Dodelson.Schneider.2013} derived the DS factor in terms of $\ns$, $\np$ and $\nd$, which is given by
\begin{align}
    f_{\rm DS} = 1+\frac{(\nd-\np)(\ns-\nd-2)}{(\ns-\nd-1)(\ns-\nd-4)}
    \label{eq:ds-factor-n}
\end{align}
in the Gaussian limit. 

\subsection{DS factor with a linear compression}
Here we consider the linear compression of the data with a compression matrix $B$. The quantities after compression is obtained by
\begin{align}
\begin{cases}
    d^c    &= B d \ ,\\
    t^c(p) &=Bt(p) \ ,\\
    S^c    &=BSB^T \ ,\\
    C^c    &= BCB^T \ , \\
    M^c    &= BM \ , \\
\end{cases}
\label{eq:compress}
\end{align}
where the superscript ${}^{c}$ stands for the compression. 
When we use the compressed data vector for the parameter estimate, the posterior width increases for two reasons; one is the DS effect. Even after the compression, depending on how we define the compression matrix, the noise in the compressed sample covariance matrix $S^c$ can still propagate to the parameter scatter. The other reason is the efficiency of the compression. If the compression is a suboptimal or lossy compression, then the posterior width increases because of lost information.

In this paper, we define the DS factor as a correction factor that accounts for the parameter scatter purely due to the propagation of the noise in the sample covariance. Therefore, we define the DS factor with a linear compression as
\begin{align}
    f_{\rm DS} = \frac{\tr\left(X_{S^c}C^cX_{S^c}^T\right)}{\tr\left(X_{C^c}C^cX_{C^c}^T\right)} \ .
    \label{eq:ds-factor-tr-c}
\end{align}
The parameter constraint is affected by both of the DS factor and the information loss in compression, and its degradation factor is given by
\begin{align}
    f_{\sigma} = \frac{\tr\left(X_{S^c}C^cX_{S^c}^T\right)}{\tr\left(X_{C}CX_{C}^T\right)} \ . 
    \label{eq:sigma-factor-tr-c}
\end{align}
When the compression is lossless, then these two factors agree. We note that the $f_\sigma$ should {\it not} be multiplied on the sample covariance during inference, because the compressed data vector is aware of the loss of the information, and the only factor need to be multiplied to the sample covariance is $f_{\rm DS}$ \footnote{And the Anderson-Hartlap factor\citep{Anderson.Anderson.2003,Hartlap.Schneider.2007}}.

Our goal is to find a better compression matrix $B$ to minimize the degradation factor $f_\sigma$ by balancing these two competing effects. We can also rephrase this as a bias-variance trade-off problem, where the former is the systematic bias in information content due to the suboptimal compression and the latter is the variance due to the noise propagation from the original sample covariance.

\section{MOPED variants and their performance}\label{sec:moped-variants}
\subsection{Standard MOPED}\label{sec:moped}
\begin{figure*}[t]
    \centering
    \includegraphics[width=0.47\linewidth]{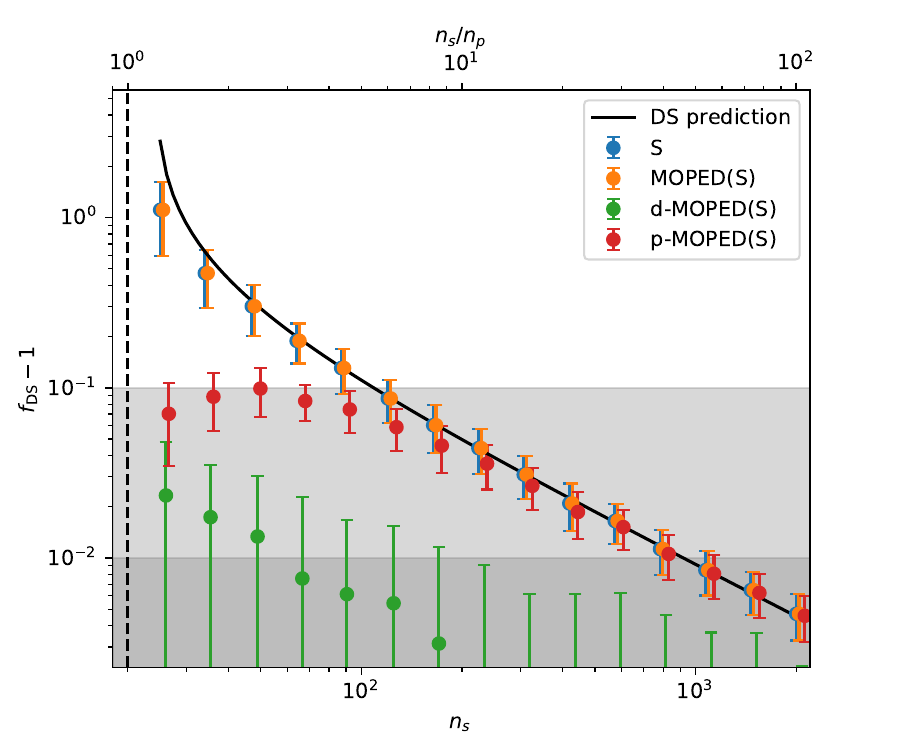}
    \includegraphics[width=0.47\linewidth]{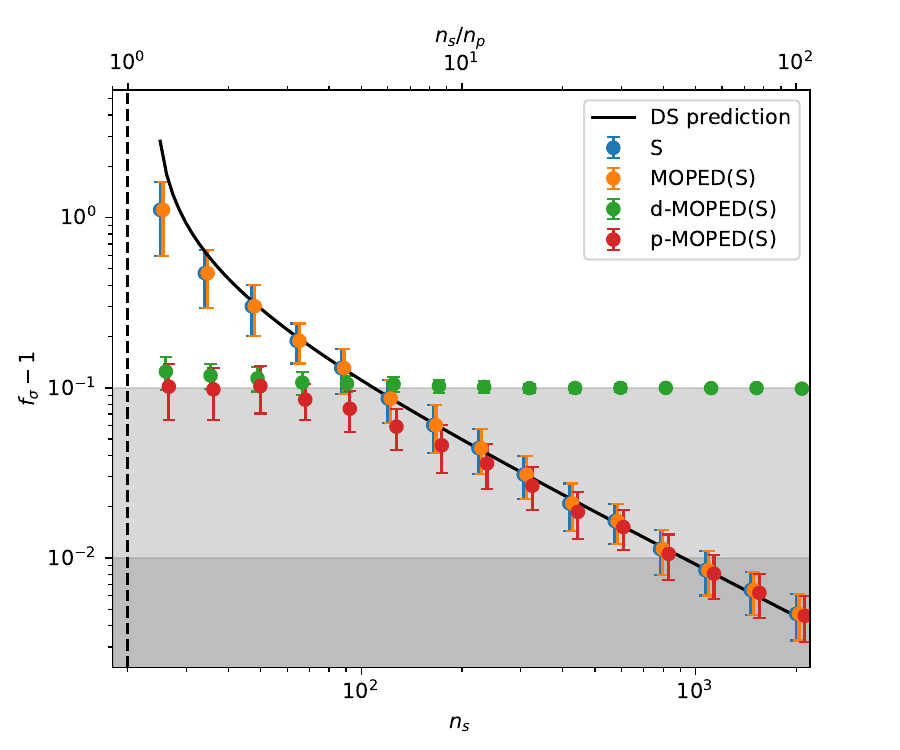}
    \caption{The performance of the various MOPED-like compression methods. 
    The left panel shows the Dodelson-Schneider (DS) factor that is needed to account for the propagation of the noise from the original sample covariance to the final posterior estimate, as a function of the number of samples $\ns$. 
    The right panel shows the degradation factor due to both  the DS effect and the information loss in data compression. 
    The solid black line shows the theoretical prediction by \citet{Dodelson.Schneider.2013} as in Eq.~(\ref{eq:ds-factor-n}). 
    The colored data points shows the result of numerical experiments, where we repeat 100 experiments; in each experiment, we draw $\ns$ samples from the true covariance, estimate the sample covariance and compression matrix, and compute the DS factor and the degradation factor by following Eqs.~(\ref{eq:ds-factor-tr-c}) and (\ref{eq:sigma-factor-tr-c}). 
    Blue points show the result without any compression. 
    The orange points show the result for the standard compression method, which needs the same correction factor and have the same degradation as no-compression case.
    The green and red points are the result with \dmoped and \pmoped respectively. 
    As designed, the degradation factor of \pmoped follows those of \dmoped and standard MOPED at low and high $\ns$ regime respectively.}
    \label{fig:corrections}
\end{figure*}
The Massively Optimized Parameter Estimation and Data compression (MOPED) \cite{Heavens.Lahav.1999}, also known as the score compression, is a linear compression that is designed to preserve Fisher information ($F_C^c = F^{\ }_C$), whose compression matrix is given by
\begin{align}
    B = M^TC^{-1}. \label{eq:moped}
\end{align}
It is common to use the ortho-normalized version of the above compression matrix by performing the Gram-Schmidt process as initially introduced in \cite{Heavens.Lahav.1999} such that the compressed covariance matrix is the identity matrix. However, the core idea and performance of compression is Eq.~(\ref{eq:moped}) and the change of basis does not affect constraining power. 

One challenging part of the standard MOPED compression is that the compression matrix is a function of the un-compressed covariance matrix $C$, which is usually not known a priori. If we naively use the sample covariance $S$ in Eq.~(\ref{eq:moped}) to obtain $B=B(S)$,
then all the noise in $S$ can still propagate to the $S^c$ through $B(S)$, leaving the same DS factor before and after compression as in Eq.~(\ref{eq:ds-factor-n}). This means that MOPED with $S$ does not reduce the effect of $f_{\rm DS}$ at all, despite the fact that after MOPED $\nd^c = \np \ll \ns$.\footnote{However, it does help to reduce the Anderson-Hartlap factor because the dimension of data vector can be significantly reduced by the compression.} 
We leave the mathematical proof of this fact for Appendix~\ref{sec:proof-ds-moped}. Furthermore, since it is a lossless compression, $\tr (X_{C^c}^{\ } C^cX_{C^c}^T) = \tr (X^{\ }_CCX_C^T)$ therefore $f_{\sigma} = f_{\rm DS}$.

One may wonder whether the noise propagation is alleviated by using two independent sample covariance estimates for compression in Eq.~(\ref{eq:moped}) and the compressed covariance estimate $S^c$ in Eq.~(\ref{eq:compress}).
This approach is commonly adopted in simulation-based inference to avoid the potential correlation of compression and inference in analogy to the train-and-validation split. However, as proved in Appendix~\ref{sec:proof-ds-moped}, the DS factor of compressed data originates from the sample covariance used for compression in Eq.~(\ref{eq:moped}), and therefore the DS factor cannot be removed as long as we use the sample covariance for compression. 

In order to explicitly see this effect, we repeated 100 experiments; in each experiment, we draw $\ns$ samples from the true covariance, estimate the sample covariance and the compression matrix, and computed the DS correction factor and the degradation factor by following Eqs~(\ref{eq:ds-factor-tr-c}) and (\ref{eq:sigma-factor-tr-c}). 
We use the example defined in Appendix~\ref{sec:example4} for the true covariance and response matrix with $(\nd,\np)=(20,2)$. 
Fig.~\ref{fig:corrections} shows the correction factors needed for the standard MOPED compression. 
We see that the correction factor for the standard MOPED compression is the same as that without compression
and follows the prediction of Eq.~(\ref{eq:ds-factor-n}). Though the compression is lossless, MOPED completely fails to deal with the situation when $\ns$ is similar to the original $\nd$ all while reducing $\nd$ to $\np$.

\subsection{Diagonal MOPED}\label{sec:dmoped}
The idea of diagonal MOPED (\dmoped) is to remove the noisy elements of the sample covariance just by simply throwing away the off-diagonal elements from the sample covariance, and is also presented in \citet{Homer.Gruen.2024}. We define the diagonalized sample covariance as 
\begin{align}
    S^d_{ij} \equiv
    \begin{cases}
        S_{ii}=\sigma_i^2 & (i=j) \\
        0 & (i\neq j)
    \end{cases}
    \label{eq:Sdiag}
\end{align}
where $S_{ii}$ is the sample variance of the $i$-th data element.
We use this diagonalized sample covariance in Eq.~(\ref{eq:moped}) to estimate the compression matrix $B$, while use the full sample covariance to obtain the compressed sample covariance in Eq.~(\ref{eq:compress}). 
The use of diagonalized sample covariance largely reduces the propagation of noise from off-diagonal elements, and therefore reduces the DS factor $f_{\rm DS}$ significantly. 
However, \dmoped neglects the cross-correlation of different data vector elements and therefore leads to suboptimal or lossy compression. Especially when the true covariance has large off-diagonal elements,  \dmoped compression can be inefficient and the degradation factor $f_\sigma$ becomes large.

Fig.~\ref{fig:corrections}, we show the performance of \dmoped. As in the last section, we use the same toy example with the same data and parameter dimensions. On the left panel, we observe that the DS factor is significantly reduced by \dmoped as expected. 
On the right panel, we observe that when the number of the simulation $\ns$ is small compared to the original data dimension $\nd$, \dmoped gives smaller degradation factors $f_\sigma$ than Eq.~(\ref{eq:ds-factor-n}). 
This is because  noise in the sample covariance is dominant when $\ns$ is small and \dmoped prevents its propagation. 
However, when the number of simulations is much larger than the data dimension $\nd$, the performance becomes worse than without  Eq.~(\ref{eq:ds-factor-n}) because the compression is suboptimal due to the missed cross-covariance structure in the diagonalized sample covariance. 
\citet{Homer.Gruen.2024} presented the result of a similar experiment in their Appendix~F, and the conclusion is consistent to our findings.

\subsection{Powered MOPED}\label{sec:pmoped}
\begin{figure*}[t]
    \centering
    \includegraphics[width=\linewidth]{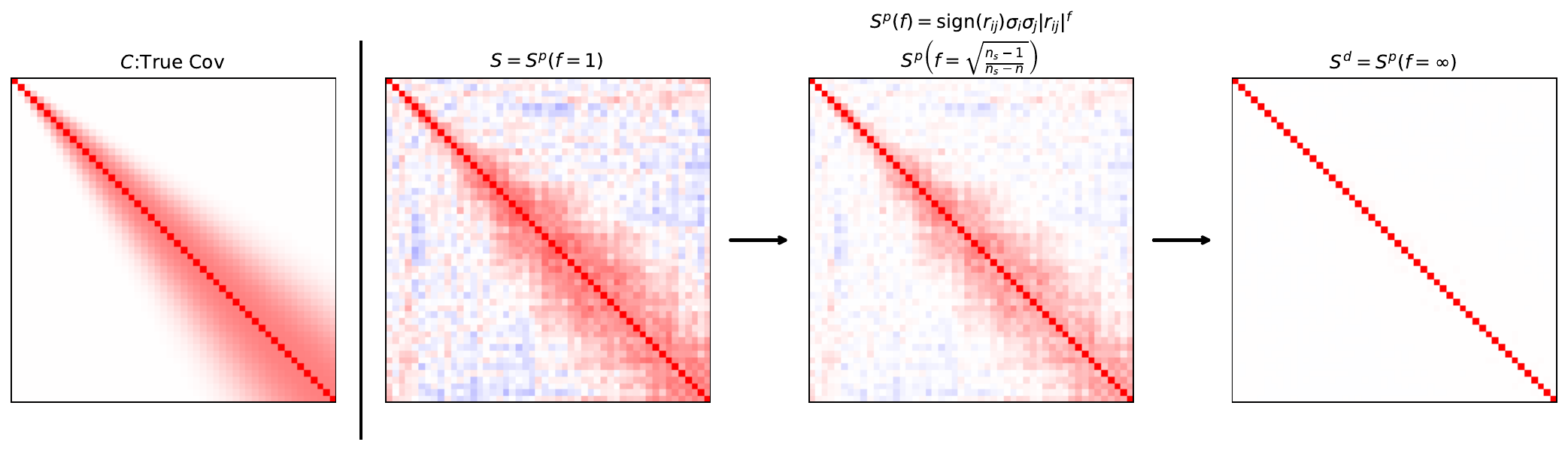}
    \caption{The idea behind the powered MOPED (\pmoped). The leftmost panel shows the true covariance matrix. A sample covariance estimate $S$ from a set of $\ns$ samples drawn from the true covariance is shown in the second leftmost panel, while the rightmost panel is the diagonalized sample covariance $S^d$ where all the off-diagonal elements are set to zero. The powered covariance matrix shown in the middle panel can be seen as a matrix between them, by adjusting the power index $f$, we suppress the size of the off-diagonal elements while keeping the structure of the cross-covariance. In this paper, we use $f=[(\ns-1)/(\ns-\nd)]^{1/2}$. The red/white/blue color indicates positive/zero/negative cross-covariance.}
    \label{fig:Sp}
\end{figure*}
The powered MOPED (\pmoped) is motivated by the above observation of the two competing effects, the noise variance and lossy compression due to diagonalized sample covariance. 
We introduce a powered sample covariance with power index $f$ defined by
\begin{align}
    S^p_{ij}(f) \equiv \sign{r^{cc}_{ij}} \sigma_i\sigma_j \left|r^{cc}_{ij}\right|^{f}
    = \left|r^{cc}_{ij}\right|^{f-1} S_{ij}\ ,
\end{align}
where $r^{cc}$ is the correlation coefficient matrix of the sample covariance and $f$ is the power index that we will adjust later to reduce the noise propagation. Note that this is an element-wise operation that preserves diagonal elements while suppressing the size of the off-diagonal elements. 
The powered sample covariance includes two limiting cases; $f=1$ corresponds to the sample covariance (Eq.~(\ref{eq:S})) and $f\rightarrow\infty$ corresponds to the diagonalized sample covariance (Eq.~(\ref{eq:Sdiag})).
For \pmoped to work as like \dmoped and standard MOPED at low and high $\ns$, respectively, we should design the dependence of $f$ on $\ns$ to have $f\rightarrow\infty$ for $\ns\sim\nd$ and $f\rightarrow1$ for $\ns\rightarrow\infty$. We use $f=[(\ns-1)/(\ns-\nd)]^{1/2}$ in this paper. 
We do not explore the further optimal choice of the exponent $f$, while readers can see e.g. \citet{Vishny.Hodyss.2024}, where authors explored a similar method to reduce the noise in the sample covariance, where they adopted the optimization of the exponent $f$ so that the amount of the suppressed noise is just below the typical noise level (See Eq. (30) of their paper).

The choice of power law as opposed to any other transformation (such as affine or sigmoid) is due to its simplicity and ability to penalize small values of $r^{cc}$ more. A $r^{cc}$ of $0.1$ is much more likely to be spurious based on limited $\ns$ compared to a $r^{cc}$ of $0.9$. So, it is desirable to have a simple transformation that brings small values to $0$ faster than it does larger values.  

Fig.~\ref{fig:Sp} visualizes the idea of the powered sample covariance. Here, we use an example model defined in Appendix~\ref{sec:example4}, and $(\nd,\np,\ns)=(50,2,100)$. We see that the off-diagonal elements are suppressed, while the structure of cross-covariance is still preserved. 

In Fig.~\ref{fig:corrections}, we show the performance of \pmoped.
The DS factor due to \pmoped shown in the left panel is always smaller than that of the standard MOPED and converges to that in the limit of $\ns\rightarrow\infty$. In the right panel, we observe that the degradation factor of \dmoped behaves similarly to \dmoped and standard MOPED in the low and high $\ns$ regime, respectively, and it smoothly transits the performance between these two limiting cases, as we designed the powered sample covariance. Moreover, the performance is always better than that of standard MOPED and \dmoped in any $\ns$.

In Appendix~\ref{sec:examples-performance}, we also show the performances of MOPED-like compressions on different toy examples. Except for the case where the true covariance is identity matrix, we see the same as in Fig.~\ref{fig:corrections}; \dmoped works better in low $\ns$ regime because the noise variance is large, but the standard MOPED works better at high $\ns$ because the sample covariance captures the true covariance structure well. We find that the $\ns$ value where the standard MOPED starts to work better than \dmoped varies depending on the covariance (it would also depend on the response matrix). This makes it difficult for the user to determine which of \dmoped and the standard MOPED to use, because the true covariance and the pivotal $\ns$ value is not known a priori. However, \pmoped naturally transits the performance in these two limits, and the user can safely use it for any $\ns$.

\section{Application: \\HSC weak lens cosmology}\label{sec:application}
\begin{figure*}[t]
    \centering
    \includegraphics[width=0.25\linewidth]{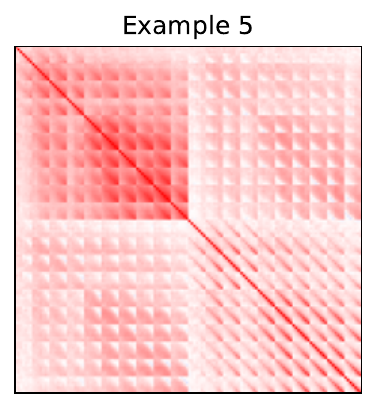}
    \hspace{-1mm}
    \includegraphics[width=0.35\linewidth]{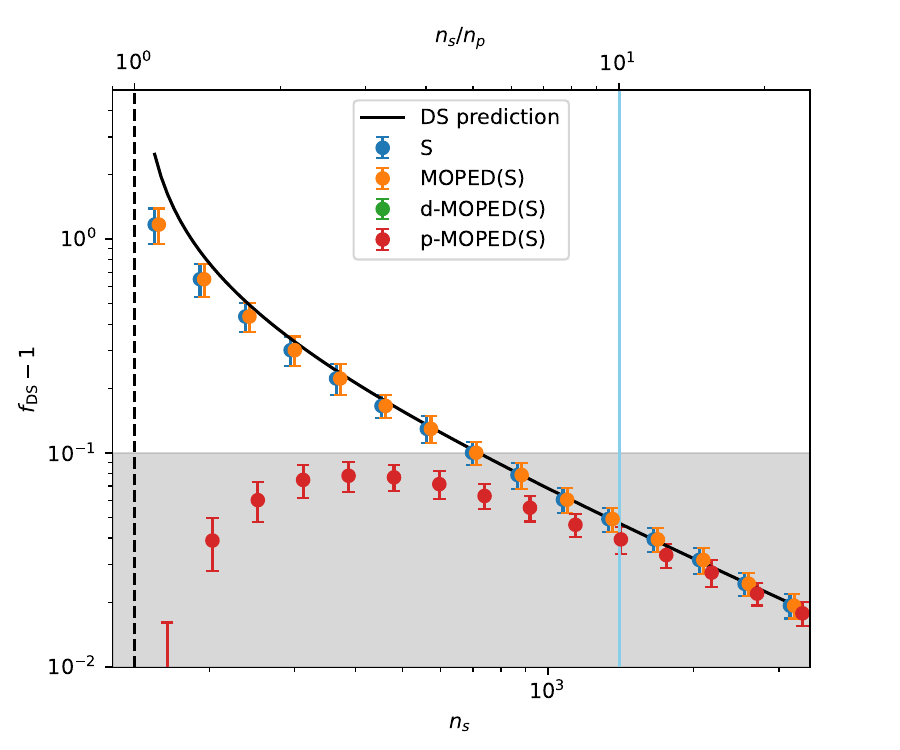}
    \hspace{-6mm}
    \includegraphics[width=0.35\linewidth]{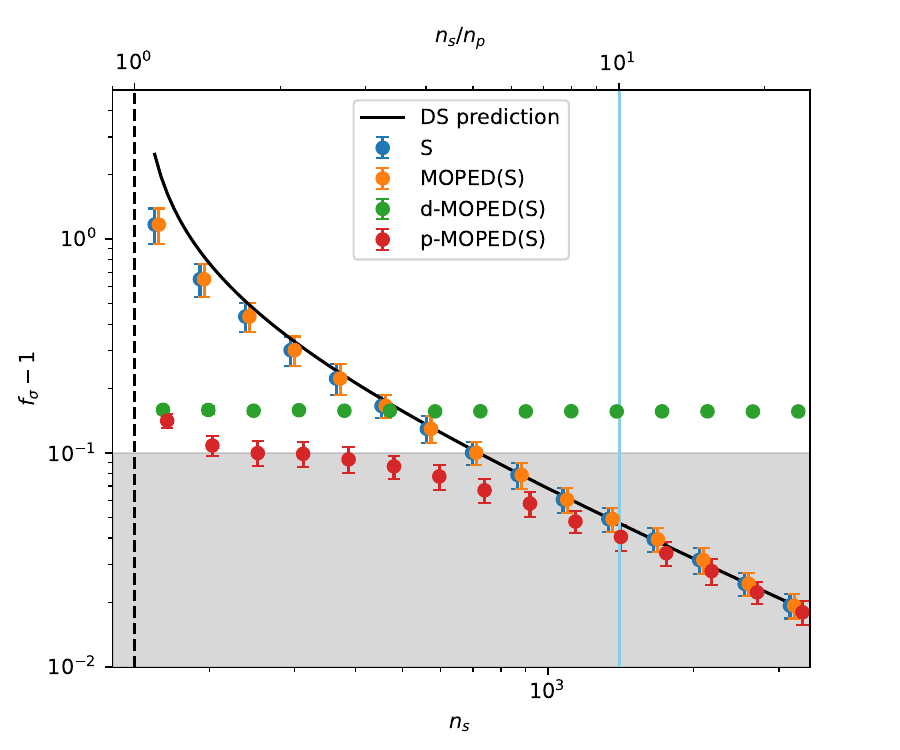}
    \includegraphics[width=0.25\linewidth]{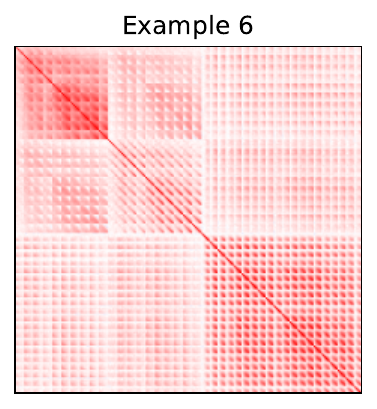}
    \hspace{-1mm}
    \includegraphics[width=0.35\linewidth]{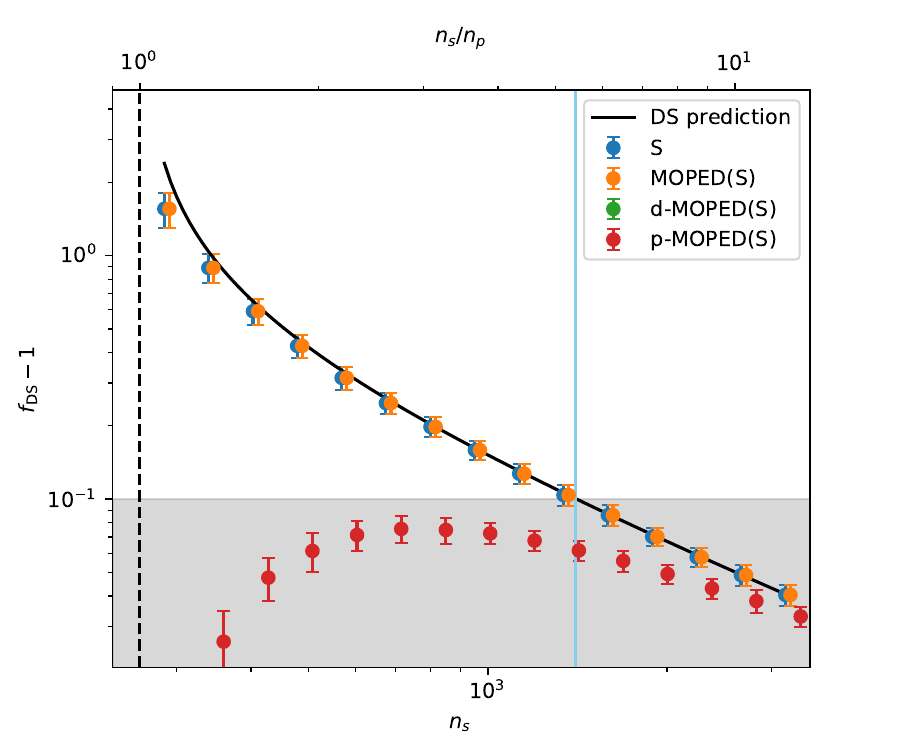}
    \hspace{-6mm}
    \includegraphics[width=0.35\linewidth]{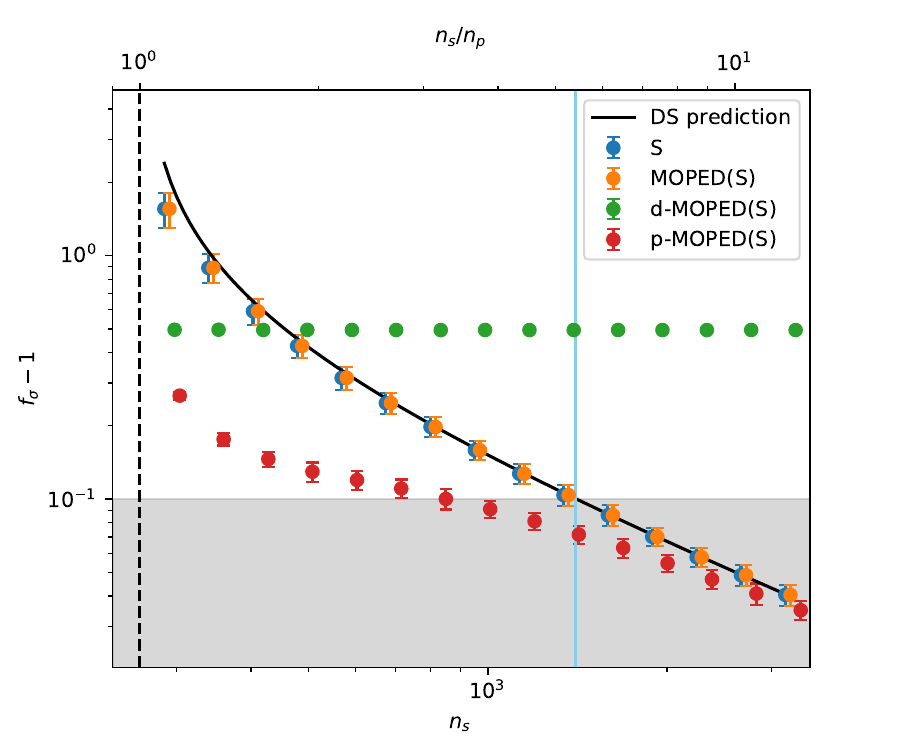}
    \caption{The result of the MOPED compression methods applied to 
    Subaru Hyper Suprime-Cam Year 3 (HSC-Y3) likelihood. 
    {\it Top:} We use the likelihood of the cosmic-shear two-point correlation function analysis from \citet{Li.Wang.2023}. The leftmost panel shows the correlation coefficient of the covariance matrix obtained from 1404 HSC-Y3 simulations, which we assume to be a true covariance. The middle and right panels are the same figure as Fig.~\ref{fig:corrections}, but with the HSC-Y3 covariance and the response matrix which is obtained from the partial derivative of the theoretical model at the best-fit model parameters. The dimensions are $(\nd,\np)=(140, 19)$. The number of the simulations (1404) are indicated by the vertical blue line in the middle and right panels. In the middle panels, the green data points for \dmoped are below the plotting range.
    {\it Bottom:} The result for the joint analysis of the 2+3PCF analysis of HSC-Y3 cosmic shear data (Sugiyama et al., in prep.). Here the dimensions are $(\nd,\np)=(300, 19)$.
    }
    \label{fig:hsc-application}
\end{figure*}
We apply the MOPED-like compressions developed in this paper to more realistic problems. Here, we consider the Subaru Hyper Suprime-Cam Year 3 (HSC-Y3) weak lensing cosmology analyses. More specifically, we use the likelihood of the two-point correlation function (2PCF) of cosmic shear analysis \cite{Li.Wang.2023}. In the paper, the covariance is estimated from $\ns=1404$ simulation data, which we assume to be a true covariance matrix in this paper. The theoretical model has 19 input parameters, which includes the cosmological, astrophysical and observational parameters. The model predicts the two-point correlation function of the cosmic shear as a function of input parameters, $t(p)$, whose dimension is $\nd=140$. The true model prediction is a nonlinear function of the input parameters, while we assume it to be a linear function $t(p)=Mp$. We compute the response matrix as the partial derivative of the model prediction with respect to the input parameters at the best-fit model parameters, $M=\partial t/\partial p$. We will leave further detail of the HSC-Y3 likelihood to \citet{Li.Wang.2023}. 

The upper panels of Fig.~\ref{fig:hsc-application} show the result of the HSC application. On the leftmost panel, we show the correlation coefficient of the measured covariance. 
The covariance matrix has non-zero off-diagonal elements, whose structure is more complicated than the toy examples we consider in the previous section and in Appendix~\ref{sec:examples}. From the degradation factor plot, we find that \pmoped provides the better compression for any $\ns$ than other MOPED-like compressions, as we have seen in the toy examples. At $\ns=1404$, the degradation factor of \pmoped is 
slightly better than the standard MOPED, and significantly better than \dmoped.

As another application, we consider the joint analysis of the 2PCF and 3PCF of the HSC-Y3 data (Sugiyama et al., in prep.). The 3PCF is one of the higher-order statistics that captures the non-Gaussian feature of the weak lensing shear field. We measured the 3PCF from the same set of 1404 simulations that we used in the HSC-Y3 2PCF analysis. Because the 3PCF is a high dimensional data vector \footnote{This is because 3PCF is a function of the triangle shape, and therefore, if we bin the triangles into 20 side length bins and 30 opening angle bins, then the 3PCF becomes a $20\times20\times30=12000$ dimensional data vector per tomographic bin. 
}, we compress it to the so-called mass skewness signal. The resultant mass skewness signal has 160 elements. Combined together with the 2PCF, the dimension of the total data vector is $\nd=300$.

The bottom panels of Fig~\ref{fig:hsc-application} show the result of the MOPED-like compressions for the 2+3PCF analysis. In the plot of the degradation factor, we find that the performance of \pmoped ($f_\sigma=6\%$) is better than the standard MOPED ($f_\sigma=10\%$) at $\ns=1404$, which suggests the use of \pmoped rather than the standard MOPED.

\section{Summary and outlook}\label{sec:summary}

In this paper, we addressed the challenge of noise propagation from sample covariance estimates to cosmological parameter inference, particularly in the context of high-dimensional data vectors and limited simulation budgets. While the MOPED algorithm offers an attractive approach for lossless linear compression that preserves Fisher information, we demonstrated that it does not mitigate the Dodelson-Schneider effect when the compression matrix is constructed using noisy sample covariance estimates.

To overcome this limitation, we proposed and tested two alternative compression strategies: diagonal MOPED (\dmoped), which discards off-diagonal elements of the sample covariance to suppress noise at the cost of some information loss, and powered MOPED (\pmoped), which introduces a tunable interpolation between standard and diagonal approaches via a power-law modification of the correlation matrix. Through extensive tests on toy models and realistic applications to Subaru HSC-Y3 weak lensing data, we showed that \pmoped consistently achieves better performance across different regimes of simulation number $n_{\rm s}$, smoothly interpolating between the low- and high-
$n_{\rm s}$ limits without requiring prior knowledge of the optimal strategy.

In realistic cosmological settings, where the data vector dimension is large and simulations are expensive, our method offers a practical and robust alternative to standard MOPED. For example, in the HSC-Y3 analysis of 2PCF and 3PCF, \pmoped yields lower degradation factors than both standard and diagonal MOPED, demonstrating its utility in high-dimensional joint analyses.

The compression method \pmoped proposed in this work is not limited to  cosmological problems, and can be applied in any scientific problem where the likelihood is Gaussian and analytical theory is available. 

Looking forward, several avenues merit exploration. First, a more systematic optimization of the power index $f$ in \pmoped could further enhance performance and generality. Second, understanding and mitigating noise propagation in other data compression techniques relying on sample covariance, such as canonical correlation analysis (CCA)~\citep{Park.Jain.2024}, could be useful for other problem settings. 
Since CCA also relies on empirical estimates of covariance matrices and response, it may suffer from similar noise. Furthermore, in simulation-based inference (SBI), both the compressed summary statistics and the mapping between parameters and data (analogous to the response matrix in our setting) must be learned from a finite number of simulations. This learning process is also susceptible to noise. Therefore, developing noise-suppression techniques like the powered compression method proposed here may play an essential role in enhancing the robustness and efficiency of SBI pipelines under realistic constraints on simulation numbers.

Our results emphasize the importance of balancing information retention and noise suppression in designing compression schemes. By explicitly controlling this trade-off, powered MOPED provides a robust compression strategy that enhances the fidelity of cosmological inference when facing practical limitations on simulation resources.

\section*{Acknowledgment}
We thank Bhuvnesh Jain, Marco Gatti, Mike Jarvis, and Rafael Gomes for useful discussion and comments on the draft. SS is supported by the JSPS Overseas Research Fellowships.

\bibliography{refs}

\appendix

\section{Proof of the invariance of DS factor with standard MOPED compression}\label{sec:proof-ds-moped}
Here we will prove that the DS factor is equivalent even after compression, when we apply the standard MOPED compression using the sample covariance to define the compression matrix.
Specifically, we consider a case where two different sets of samples are used to construct the MOPED compression matrix and to estimate the sample covariance for compressed data.
In the standard MOPED scenario, we construct the compression matrix as
\begin{align}
    B = M^{T}S_1^{-1}\ .
\end{align}
Here, $S_1$ is the sample covariance estimated from a first set of samples.
With this definition of compression matrix, we obtain the explicit expression for the matrices associated with the compressed data vector.
\begin{align}
     S^{\rm c} &\equiv BS_2B^{T} = M^{T}S^{-1}_1S_2S^{-1}_1M \nonumber \\
     C^{\rm c} &\equiv BCB^{T} = M^{T}S^{-1}_1CS^{-1}_1M \nonumber  \\
     M^{\rm c} &\equiv BM  = M^{T}S^{-1}_1M = F_{S_1} \ .
\end{align}
Here, $S_2$ is the sample covariance estimated from a second set of samples. 
Note that $B$ has units of $[{\rm data }]/[{\rm parameter }]$. So, compressed data vectors have units of $1/[{\rm parameter }]$. The Fisher matrix with compressed sample covariance matrix is
\begin{align}
    F_{S^{\rm c}} 
    &\equiv (M^{\rm c})^{T} (S^{\rm c})^{-1} (M^{\rm c}) \nonumber\\
    & = F_{S_1} (S^{\rm c})^{-1} F_{S_1}
\end{align}
and the response matrix of parameter with respect to the compressed data is
\begin{align}
    X_{S^{\rm c}}
    &\equiv F_{S^{\rm c}}^{-1} (M^{\rm c})^{T} (S^{\rm c})^{-1} \nonumber\\
    &=F_{S_1}^{-1}S^cF_{S_1}^{-1} F_{S_1} (S^{\rm c})^{-1} \nonumber \\
    &=F_{S_1}^{-1} \ .
\end{align}
By substituting the above two expressions to the numerator of the DS factor for the compressed matrix, we obtain
\begin{align}
    X_{S^{c}} C^{c} X_{S^{c}}^{T} 
    &= F_{S_1}^{-1}M^{T} S_1^{-1} C S_1^{-1} M (F_{S_1}^{-1})^{T} \nonumber \\
    &= X_{S_1} C X_{S_1}^{T} \ , 
\end{align}
which shows the equivalence of the numerators of the DS factor before and after the compression. In a similar way, we can prove the equivalence for the denominator. This means that the noise propagated to the compressed data is only from the sample covariance used to define the compression matrix $S_1$ and not from the sample covariance used to quantify the covariance of the compressed data vector $S_2$.

\section{Collection of toy examples}\label{sec:examples}
\begin{figure*}[t]
    \centering
    \includegraphics[width=0.22\linewidth]{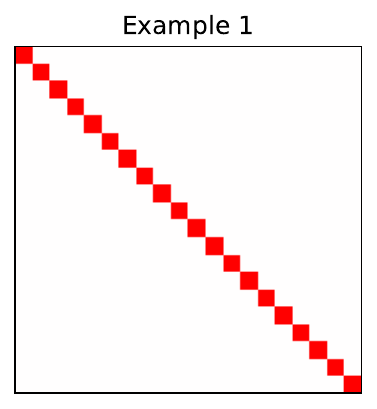}
    \includegraphics[width=0.22\linewidth]{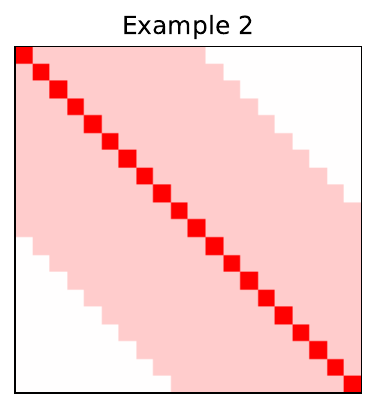}
    \includegraphics[width=0.22\linewidth]{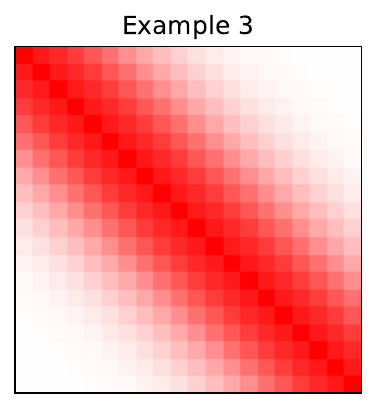}
    \includegraphics[width=0.22\linewidth]{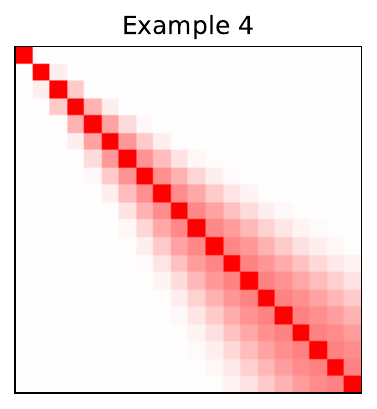}
    \caption{The covariance matrices of the toy examples used in this paper. Here we show the covariance matrices with dimension $\nd=20$. From left to right the complexity of the covariance structure increases; the covariance of Example~1 is the identity matrix for which \dmoped works extremely well. From Example~2, we add various off-diagonal elements, to test the performance of various MOPED-like compressions, especially \dmoped and \pmoped. Example~2 has a constant off-diagonal element up to 10 neighboring elements. Example~3 has the exponentially decaying off-diagonal elements, whose decay length is set to be constant $\Delta_{ij}=1$. Example~4 is similar to Example~3, but its decaying length increases $\Delta_{ij}=\min(i,j)/20$.}
    \label{fig:example-models}
\end{figure*}
In this section, we present a collection of the true covariance that we use as toy example problem to test the performance of the MOPED variants in this paper.
Throughout this paper, we use the response matrix defined by
\begin{align}
    M_{ij} = 
    \begin{cases}
        1, & i=j~({\rm mod}~\np)\\
        0, & \text{otherwise}
    \end{cases},
\end{align}
unless otherwise noted.

\subsection{Example 1}\label{sec:example1}
This is the simplest example, where the covariance matrix is an identity matrix, $C=I$. In this example, the true covariance is diagonal, and therefore \dmoped works very well.

\subsection{Example 2}\label{sec:example2}
We use the same response matrix as Example~1
\begin{align}
    C_{ij} = 
    \begin{cases}
        1, & i=j\\
        r, & 1<|i-j| \leq \Delta\\
        0, & \text{otherwise}
    \end{cases}
\end{align}
where we set $\Delta=10$. This sets constant value $r$ on off-diagonal for the neighboring elements within $\Delta$.

\subsection{Example 3}\label{sec:example3}
In this example, the covariance has the exponentially decaying off-diagonal elements.
\begin{align}
    C_{ij} = 
    A_{ij}\exp\left(-\frac{(i-j)^2}{2 \Delta_{ij}^2}\right)
    \label{eq:exp-covariance}
\end{align}
where the amplitude is
\begin{align}
    A_{ij} = 
    \begin{cases}
        1, & i\neq j \\
        1/\Delta m, & i =j
    \end{cases}
\end{align}
We use the constant decay length $\Delta_{ij}=1$, and $\Delta m=1.1$. We introduce $\Delta m>1$ to avoid the accidental low-rank sample covariance when $\ns/\nd$ is extremely small.

\subsection{Example 4}\label{sec:example4}

In this example, we use the same covariance form as in Eq.~(\ref{eq:exp-covariance}), but we use a varying decay length $\Delta_{ij} = \min(i, j)/20$.

\

\section{Performance tests on the toy examples}\label{sec:examples-performance}

\begin{figure*}[t]
    \centering
    \includegraphics[width=0.26\linewidth]{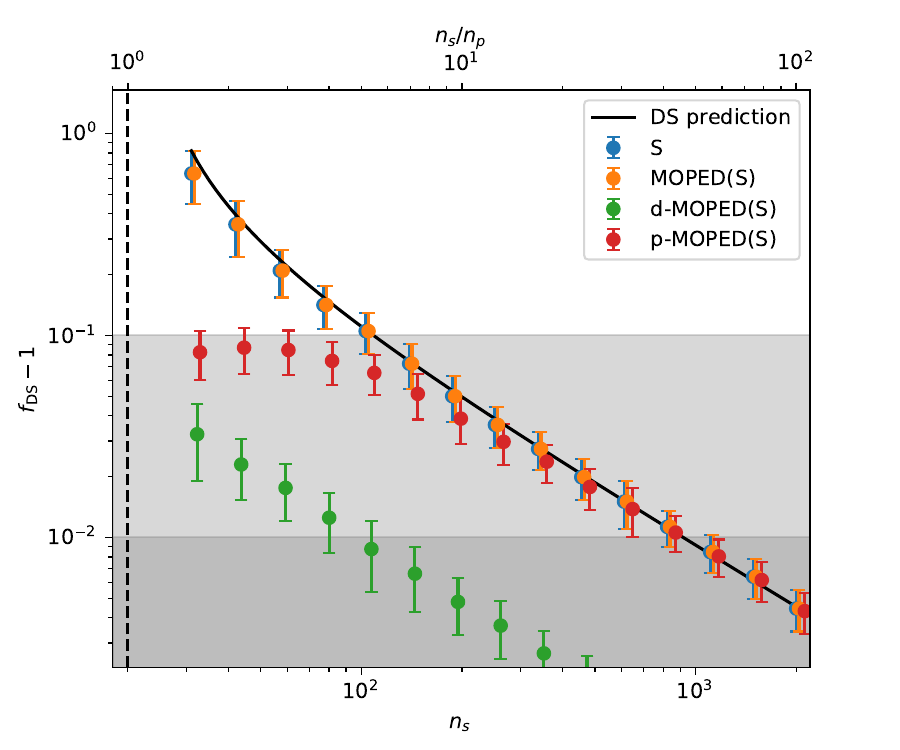}
    \hspace{-5mm}
    \includegraphics[width=0.26\linewidth]{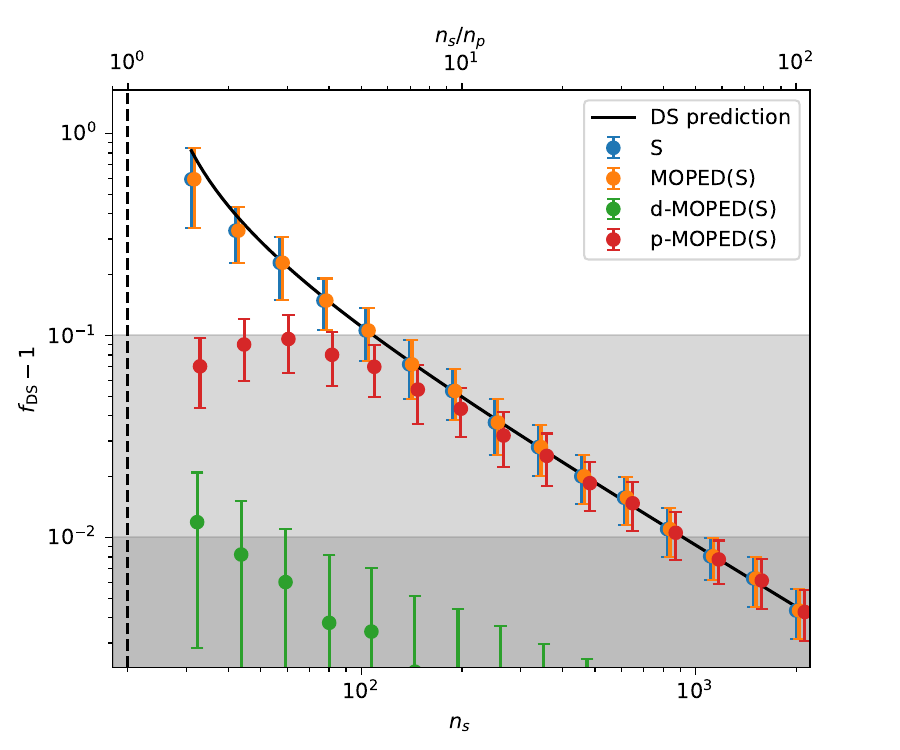}
    \hspace{-5mm}
    \includegraphics[width=0.26\linewidth]{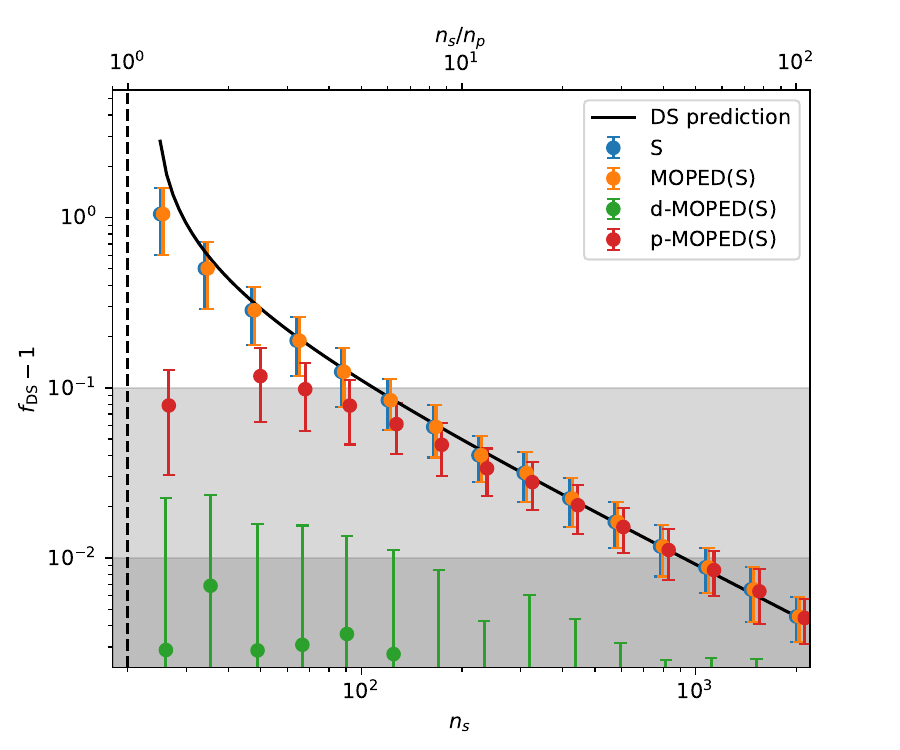}
    \hspace{-5mm}
    \includegraphics[width=0.26\linewidth]{example4_fds.pdf}
    \includegraphics[width=0.26\linewidth]{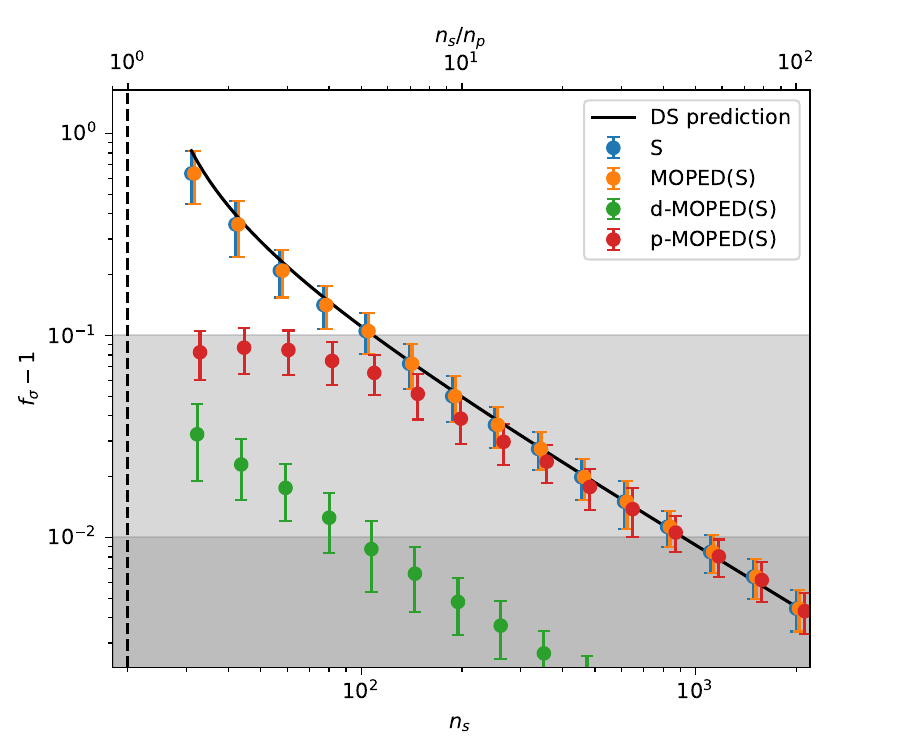}
    \hspace{-5mm}
    \includegraphics[width=0.26\linewidth]{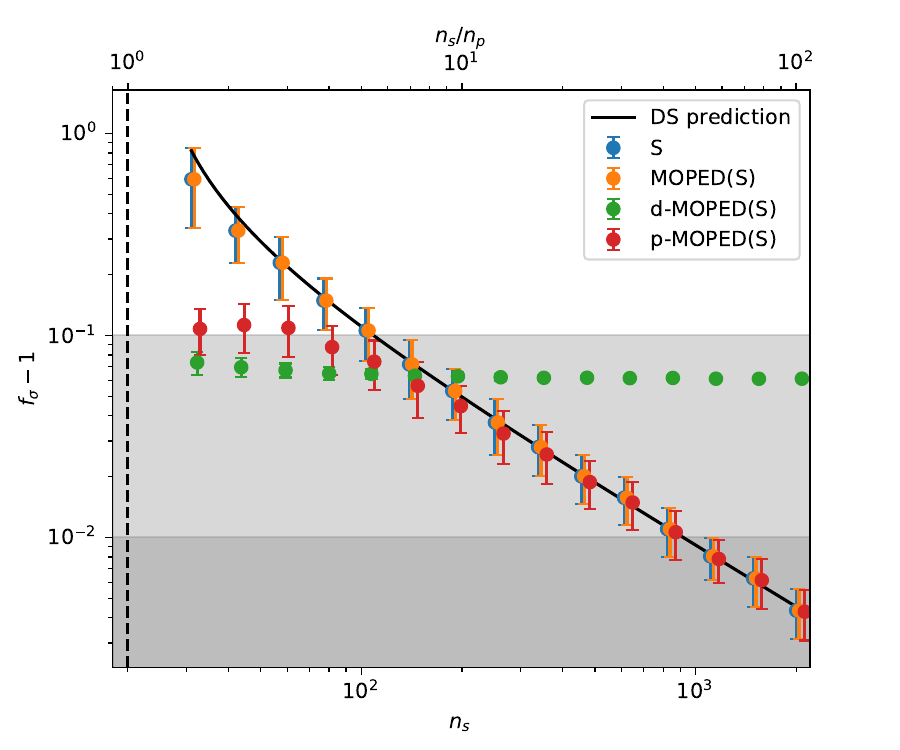}
    \hspace{-5mm}
    \includegraphics[width=0.26\linewidth]{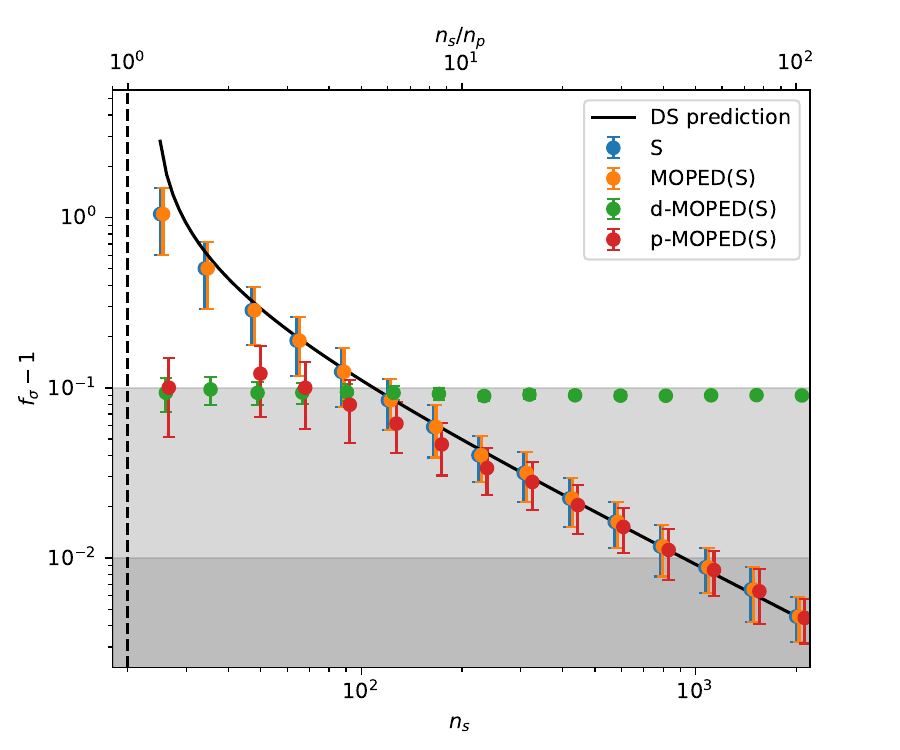}
    \hspace{-5mm}
    \includegraphics[width=0.26\linewidth]{example4_fsigma.pdf}
    \caption{The performance of MOPED variants discussed in this paper (standard MOPED, \dmoped, \pmoped). The top panels show the Dodelson-Schneider factor, $f_{\rm DS}$ defined in Eq.~(\ref{eq:ds-factor-tr-c}) and the bottom panels show the degradation factor, $f_{\sigma}$ defined in Eq.~(\ref{eq:sigma-factor-tr-c}).  From left to right, the results for Example~1,2,3, and 4 are shown. The color codes are the same as in Fig.~\ref{fig:corrections}. The rightmost panels are the identical to the two panels in Fig.~\ref{fig:corrections}, but we show it for the ease of comparison to the results on other examples.}
    \label{fig:performance-examples}
\end{figure*}
In this section, we show the results of the MOPED variants on the examples introduced in the previous section.

\end{document}